\documentclass[twocolumn]{aastex702}
\usepackage{amsmath,amssymb}
\shorttitle{Estimator Precision and Selection}
\shortauthors{Meng et al.}
\begin{document}

\title{Estimator Precision and Selection for Up-the-Ramp Sampling with a High-Read-Noise InGaAs Camera}
\correspondingauthor{Fang-Yu Xu}

\author[0009-0004-9119-6543]{Wei-Jie Meng}
\affiliation{Yunnan Observatories, Chinese Academy of Sciences, Kunming 650216, China}
\affiliation{Yunnan Key Laboratory of Solar Physics and Space Science, Kunming 650216, China}
\affiliation{Yunnan Engineering Center for Applied Astronomical Technology, Kunming 650216, China}
\affiliation{University of Chinese Academy of Sciences, Beijing 100049, China}
\email{mengweijie@ynao.ac.cn}

\author[0009-0003-5995-8125]{Fang-Yu Xu}
\affiliation{Yunnan Observatories, Chinese Academy of Sciences, Kunming 650216, China}
\affiliation{Yunnan Key Laboratory of Solar Physics and Space Science, Kunming 650216, China}
\affiliation{Yunnan Engineering Center for Applied Astronomical Technology, Kunming 650216, China}
\email[show]{{\fontencoding{T1}\selectfont xu\textunderscore fangyu@ynao.ac.cn}}

\author[0000-0001-7575-5449]{Zhen-Yu Jin}
\affiliation{Yunnan Observatories, Chinese Academy of Sciences, Kunming 650216, China}
\affiliation{Yunnan Key Laboratory of Solar Physics and Space Science, Kunming 650216, China}
\affiliation{Yunnan Engineering Center for Applied Astronomical Technology, Kunming 650216, China}
\email{kim@ynao.ac.cn}

\author[0000-0001-5546-1021]{Guang-Qian Liu}
\affiliation{Yunnan Observatories, Chinese Academy of Sciences, Kunming 650216, China}
\affiliation{Yunnan Key Laboratory of Solar Physics and Space Science, Kunming 650216, China}
\affiliation{Yunnan Engineering Center for Applied Astronomical Technology, Kunming 650216, China}
\email{lgq@ynao.ac.cn}

\begin{abstract}
The noise properties of an InGaAs camera vary with gain, requiring a trade-off between implementation complexity and precision. Using generalized least squares (GLS) with known covariance as the precision benchmark, we examine the applicability and precision penalties of ordinary least squares (OLS) and correlated double sampling (CDS) at different noise operating points. We define $\eta=fT/\sigma_\mathrm{R}^2$, where $f$ is the total pixel count rate, $T$ the fitting span, and $\sigma_\mathrm{R}$ the single-read noise. For constant flux, uniform sampling, and independent white read noise plus Poisson noise, established finite-sampling variance expressions give equal OLS and CDS variances at $\eta=10$ for $N\ge4$ reads; OLS is favored below this boundary and CDS above it. Analytical and Monte Carlo comparisons for three gain modes quantify the precision loss of the better simplified estimator relative to GLS. In high-gain on-sky data, choosing OLS at 20\,s and CDS at 90\,s increases the background residual standard deviation by 3.55\% and 3.01\%, respectively, relative to two-iteration feasible GLS (FGLS). Photometry of 21 common sources does not clearly resolve most estimator differences. The observed OLS--CDS crossover occurs earlier than the ideal prediction, indicating that temporal covariance beyond the ideal model also affects the selection boundary. Observational validation is limited to high gain; medium- and low-gain choices rely on theory and simulations using calibrated parameters. OLS or CDS can replace GLS with a small precision penalty at suitable operating points. The ideal criterion provides an initial selection rule, whose practical boundary requires validation against observed temporal covariance.
\end{abstract}


\section{Introduction}
\label{sect:intro}

Nondestructive readout (NDRO) allows the accumulated pixel signal to be sampled repeatedly within an integration, reducing the impact of read noise in infrared detectors. Correlated double sampling (CDS) takes the difference between the first and last reads to remove their common offset and reset noise. Fowler sampling combines multiple reads at each end of an integration, whereas up-the-ramp sampling records the signal growth throughout the integration to estimate the count rate \citep{Garnett1993,Fowler1991}. Reads within a ramp share the previously accumulated photon signal, so their Poisson noise is temporally correlated. The precision gained from additional reads therefore depends on the relative contributions of read and Poisson noise and on the adopted weights \citep{Rauscher2007}.

Ordinary least squares (OLS) estimates the slope by minimizing equally weighted squared residuals and is straightforward to implement. Generalized least squares (GLS) constructs weights from the full covariance and achieves the minimum variance among linear unbiased estimators when that covariance is known. For ramp fitting, \citet{Fixsen2000} developed approximately optimal weights that vary with signal-to-noise ratio, \citet{Rauscher2007} derived finite-sampling variance expressions accounting for correlated photon noise between reads, and \citet{Brandt2024} exploited the covariance of read differences to obtain optimal fits with a computational cost linear in the number of read groups, while examining biases introduced by estimating the covariance from the data. These studies provide a basis for comparing estimator precision and choosing an estimator for a given noise operating point and acceptable precision loss.

InGaAs detectors have been used for ground-based near-infrared astronomy, with observational demonstrations of read-noise reduction through ramp sampling \citep{Sullivan2013}. Our prototype InGaAs camera has been delivered for use on the Sun Yat-sen University 80\,cm infrared telescope \citep{Dong2026}. Routine observations use the detector's ITR mode with correlated double sampling (CDS). We conducted dedicated observations on this telescope to assess the performance of repeated nondestructive readout within individual integrations. Under the same environmental conditions, the camera's gain modes have different read-noise and dark-background levels, potentially favoring different estimators even at the same illumination and sampling configuration. Motivated by computational efficiency and implementation complexity, we use GLS as a precision benchmark to identify operating points where OLS or CDS can achieve comparable precision and to quantify their precision penalties in actual measurements.

Using constant flux, independent white read noise, and Poisson noise as the analytical baseline, we obtain an OLS--CDS crossover criterion from established finite-sampling variance expressions for uniform sampling with $N\ge4$. We define the operating point $\eta=fT/\sigma_\mathrm{R}^2$ in terms of the total pixel count rate $f$, fitting span $T$, and single-read noise $\sigma_\mathrm{R}$. The boundary $\eta=10$ determines which of OLS and CDS has the lower variance. Calibrated parameters for three gain modes and simulations matched to the sampling configurations then quantify the precision penalty of the better of these two estimators relative to GLS.

The incident flux can also vary during ground-based observations, in which case the read weights affect both bias and precision \citep{Li2026}. Choices based on the ideal model therefore require evaluation against actual time series. We use high-gain dark and on-sky data to compare precision penalties in background measurements and source photometry, test the crossover boundary through read-prefix scans and integrating-sphere measurements at several illumination levels, and describe the observed shift using empirical temporal covariance. GLS in the theory and simulations uses known covariance; the observational products use two-iteration feasible generalized least squares (FGLS).

Section~\ref{sect:theory} presents the estimator-selection criterion. Section~\ref{sect:simulation} describes the camera operating points and validation under the ideal model. Sections~\ref{sect:observational_methods} and~\ref{sect:results} describe the observational comparisons and their results. Sections~\ref{sect:discussion} and~\ref{sect:conclusions} discuss practical selection and summarize the conclusions.

\section{Estimator Comparison and Selection}
\label{sect:theory}

Consider uniformly sampled nondestructive reads under constant flux, with independent white read noise and accumulated Poisson noise. Let $N$ denote the number of retained reads, $\Delta t$ their spacing, $T=(N-1)\Delta t$ the fitting span, and $f=f_\mathrm{s}+f_\mathrm{b}+f_\mathrm{d}$ the total pixel count rate, comprising source, sky, and dark-background contributions. The single-read noise is $\sigma_\mathrm{R}$. CDS uses the first and last reads, OLS fits a line to all reads, and GLS determines its weights from the known covariance. We adopt established noise models and estimator expressions \citep{Rauscher2007,Brandt2024}; Appendix~\ref{sect:estimator_details} gives the model definitions, variance expressions, and GLS solution.

The dimensionless operating point $\eta=fT/\sigma_\mathrm{R}^2$ is the ratio of the Poisson variance accumulated over the fitting span to the single-read noise variance. At fixed gain, background, and read spacing, $\eta=[f\Delta t/\sigma_\mathrm{R}^2](N-1)$ is proportional to the number of retained read intervals, $N-1$, or equivalently increases linearly with $T$.

Because GLS minimizes the variance among all linear unbiased estimators when the covariance is known, we express the performance of CDS and OLS as variance ratios relative to GLS.

\subsection{Relative Efficiency in the Noise Limits}
\label{sect:limiting}

Table~\ref{tab:limiting-variance-ratios} summarizes the behavior of the three estimators in the two limiting noise regimes.

\begin{table*}[!htbp]
\centering
\caption{Finite-sampling variance ratios in the limiting noise regimes ($N\ge2$), relative to GLS with known covariance.\label{tab:limiting-variance-ratios}}
\small
\begin{tabular}{lccc}
  \hline\hline\noalign{\smallskip}
  Noise Regime & GLS & OLS & CDS \\
  \hline\noalign{\smallskip}
  Read-noise dominated ($f \to 0$) & 1 & 1 & $\dfrac{N(N+1)}{6(N-1)}$ \\
  Poisson dominated ($\sigma_\mathrm{R} \to 0$) & 1 & $\dfrac{6(N^2+1)}{5N(N+1)}$ & 1 \\
  \noalign{\smallskip}\hline
\end{tabular}

\end{table*}

The limiting behavior follows directly from the weights. When read noise dominates, the covariance approaches a diagonal matrix and GLS reduces to OLS. CDS uses only two reads, and its variance is approximately $N/6$ times the GLS variance at large $N$. When Poisson noise dominates, the inter-read increments are independent and identically distributed, and GLS reduces to CDS. Because OLS does not account for the inter-read Poisson covariance in its weights, its variance is larger by a factor $6(N^2+1)/[5N(N+1)]$, approaching $6/5$ at large $N$.

\subsection{The OLS--CDS Crossover Criterion}

From the established finite-sampling variance expressions \citep{Rauscher2007} given in Appendix~\ref{sect:cds}, the OLS--CDS variance difference is
\begin{equation}
\begin{split}
 &\mathrm{Var}(\hat f_{\mathrm{OLS}})-\mathrm{Var}(\hat f_{\mathrm{CDS}})\\
 & =\frac{(N-2)(N-3)}{5N(N+1)T^2}\left(fT-10\sigma_\mathrm{R}^2\right)\,.
 \end{split}
\label{eq:ols_cds_variance_difference}
\end{equation}
For $N\ge4$, the prefactor is positive, and the variances are equal when
\begin{equation}
  \eta \equiv \frac{f\,T}{\sigma_\mathrm{R}^2} = 10\,.
  \label{eq:crossover}
\end{equation}

For uniform sampling with $N\ge4$, the crossover condition is independent of $N$; OLS and CDS are identical for $N=2$ or 3. OLS has the lower variance for $\eta<10$, and CDS for $\eta>10$. The equivalent critical count rate is $f_\mathrm{cross}=10\sigma_\mathrm{R}^2/T$. This boundary determines the relative advantage of OLS and CDS; whether the better estimator can replace GLS also depends on its precision penalty. Variance ratios relative to GLS depend on both $N$ and $\eta$. At fixed $N$, the penalty can be compared using $\eta$, whereas comparisons across sampling configurations must also account for changes in $N$.

\section{Camera Parameters and Ideal-Model Validation}
\label{sect:simulation}

We apply the selection criterion to representative camera operating points and use Monte Carlo simulations matched to the sampling configurations to check analytical variances and aperture-error propagation and to compare estimator efficiency.

\subsection{Camera Parameters and Operating Points}
\label{sect:model_parameters}

The camera supports high-, medium-, and low-gain modes (HG, MG, and LG); the on-sky tests use HG. Table~\ref{tab:detector-parameters} lists the mode parameters. The gain converts digital counts to electrons and was measured in the laboratory using photon transfer curves. Read noise and dark count rates were measured from on-site dark ramps after installation; the latter are used as the model dark-background rates.

\begin{table*}[!htbp]
\centering
\caption{Selected parameters of the three InGaAs camera gain modes.\label{tab:detector-parameters}}
\setlength{\tabcolsep}{3pt}
\small
\begin{tabular}{lccc}
  \hline\hline\noalign{\smallskip}
  Parameter & HG & MG & LG \\
  \hline\noalign{\smallskip}
  Gain (e$^-$\,ADU$^{-1}$) & 0.466 & 1.92 & 36.7 \\
  Read noise $\sigma_\mathrm{R}$ (e$^-$) & 24 & 41 & 356 \\
  Dark-background rate $f_\mathrm{d}$ (e$^-\,\mathrm{s}^{-1}\,\mathrm{pix}^{-1}$) & 36 & 43 & 32 \\
  \noalign{\smallskip}\hline
\end{tabular}

\end{table*}

The observations use multiple reads within each integration, with a nominal spacing of $\Delta t=0.75$\,s. The analytical comparisons, Monte Carlo simulations, and dark and on-sky analyses use the two configurations in Table~\ref{tab:exposure-configurations}, referred to as the 20\,s and 90\,s configurations.

\begin{table*}[!htbp]
\centering
\caption{Definitions of the two exposure configurations.\label{tab:exposure-configurations}}
\setlength{\tabcolsep}{5pt}
\small
\begin{tabular}{lccc}
  \hline\hline\noalign{\smallskip}
  Configuration & Retained Reads & Read Count $N$ & Fitting Span $T$ \\
  & & & (s) \\
  \hline\noalign{\smallskip}
  20\,s & 2--29 & 28 & 20.25 \\
  90\,s & 2--120 & 119 & 88.5 \\
  \noalign{\smallskip}\hline
\end{tabular}

\tablecomments{Each raw ramp contains 121 reads. The first and last deviate from the linear trend of the retained sequence and are discarded, leaving 119 usable reads. Both configurations use subsets of the same ramps.}
\end{table*}

Pixelwise read noise is estimated from adjacent-read differences in 59 HG dome-dark ramps and combined using a $3\sigma$-clipped mean to obtain a read-noise map. Its spatial median is approximately 24\,e$^-$, and the root mean square (RMS) over valid full-frame pixels is 29.45\,e$^-$. The representative analytical operating points and Monte Carlo comparisons use the median of 24\,e$^-$.

The HG dark data yield a dark-background rate of $f_\mathrm{d}\approx36$\,e$^-\,\mathrm{s}^{-1}\,\mathrm{pix}^{-1}$. The $J$-band sky-background rate estimated from 44 on-sky ramps is $f_\mathrm{b}\approx77$\,e$^-\,\mathrm{s}^{-1}\,\mathrm{pix}^{-1}$. We adopt this sky rate for all gain modes and add their respective dark-background rates. The total rates $f=f_\mathrm{b}+f_\mathrm{d}$ at the representative background operating points are approximately 113, 120, and 109\,e$^-\,\mathrm{s}^{-1}\,\mathrm{pix}^{-1}$ for HG, MG, and LG, respectively.

\begin{figure*}[!htbp]
   \centering
   \includegraphics[width=0.96\textwidth, angle=0]{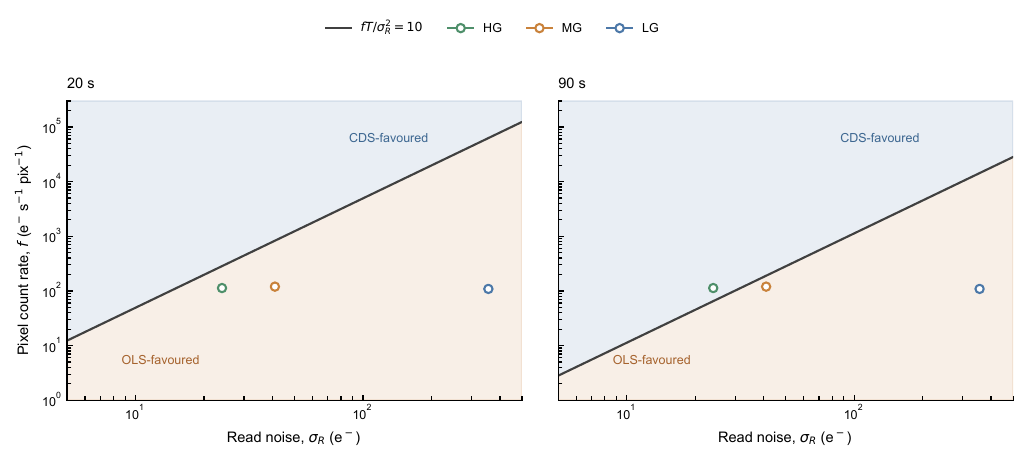}
\caption{Estimator regimes defined by the dimensionless criterion. The horizontal and vertical axes show read noise $\sigma_\mathrm{R}$ and total pixel count rate $f$. The gray boundaries indicate $fT/\sigma_\mathrm{R}^2=10$: OLS has the lower variance below each boundary, whereas CDS has the lower variance above it. Colored points mark representative background rates for the camera's HG, MG, and LG modes.}
   \label{fig:estimator-regimes}
\end{figure*}

These count rates, read-noise levels, and sampling configurations determine the representative values of $\eta$. The 20\,s and 90\,s configurations illustrate estimator selection and its precision penalty at these operating points. According to Equation~\ref{eq:crossover}, for HG, OLS is favored at 20\,s and CDS at 90\,s, whereas both configurations favor OLS in MG and LG (Figure~\ref{fig:estimator-regimes}).

\subsection{Simulation Setup}
\label{sect:sim_setup}

We generate ramps with $N$ reads following the signal model in Appendix~\ref{sect:signal_model} (Equation~\ref{eq:signal_model}), with a fixed offset $b=1000\,\mathrm{e^-}$. Each read includes independent Gaussian noise $r_i\sim\mathcal N(0,\sigma_\mathrm R^2)$. Electron increments between adjacent reads independently follow $\Delta p_j\sim\mathrm{Poisson}(f\Delta t)$ ($j=1,\ldots,N-1$). The accumulated Poisson fluctuation is $p_i=\sum_{k=1}^{i-1}(\Delta p_k-f\Delta t)$, giving total noise $n_i=r_i+p_i$.

For both sampling configurations in Table~\ref{tab:exposure-configurations}, we sample 17 logarithmically spaced total count rates $f$ from $1$ to $3\times10^5$\,e$^-\,\mathrm{s}^{-1}\,\mathrm{pix}^{-1}$. Each combination of gain, configuration, and count rate has 5000 independent simulated ramps. CDS, OLS, and GLS are applied to the same ramps, with the GLS covariance constructed from the true input count rate and read noise.

The aperture simulations use a Gaussian point-spread function (PSF) with FWHM 4\,pix, an aperture radius of 6\,pix, and a sky annulus spanning 12--18\,pix. The sky rate is 77\,e$^-\,\mathrm{s}^{-1}\,\mathrm{pix}^{-1}$, with the dark background appropriate to each gain. Total source count rates span $10^2$--$10^5$\,e$^-\,\mathrm{s}^{-1}$ at 17 logarithmically spaced points, with 5000 realizations per combination. All three estimators use the same photometric operator.

For independent pixels and a sky estimate given by the annular mean, let $\sigma_{\hat f,i,\mathrm{formal}}^2$ be the formal variance of the estimated count rate in pixel $i$. The formal variance of the net aperture count rate is
\begin{equation}
  \sigma_\mathrm{ap,formal}^2 =
  \sum_{i\in\mathrm{ap}}\sigma_{\hat f,i,\mathrm{formal}}^2
  + \left(\frac{A_\mathrm{ap}}{N_\mathrm{sky}}\right)^2
  \sum_{j\in\mathrm{sky}}\sigma_{\hat f,j,\mathrm{formal}}^2 \,.
  \label{eq:aperture_rate_uncertainty}
\end{equation}
where $A_\mathrm{ap}$ is the aperture area, $N_\mathrm{sky}$ is the number of sky-annulus pixels, and $\sigma_\mathrm{ap,formal}$ is in e$^-\,\mathrm{s}^{-1}$.

\subsection{Simulation Results}
\label{sect:sim_results}

\begin{figure*}[!htbp]
   \centering
   \includegraphics[width=0.96\textwidth, angle=0]{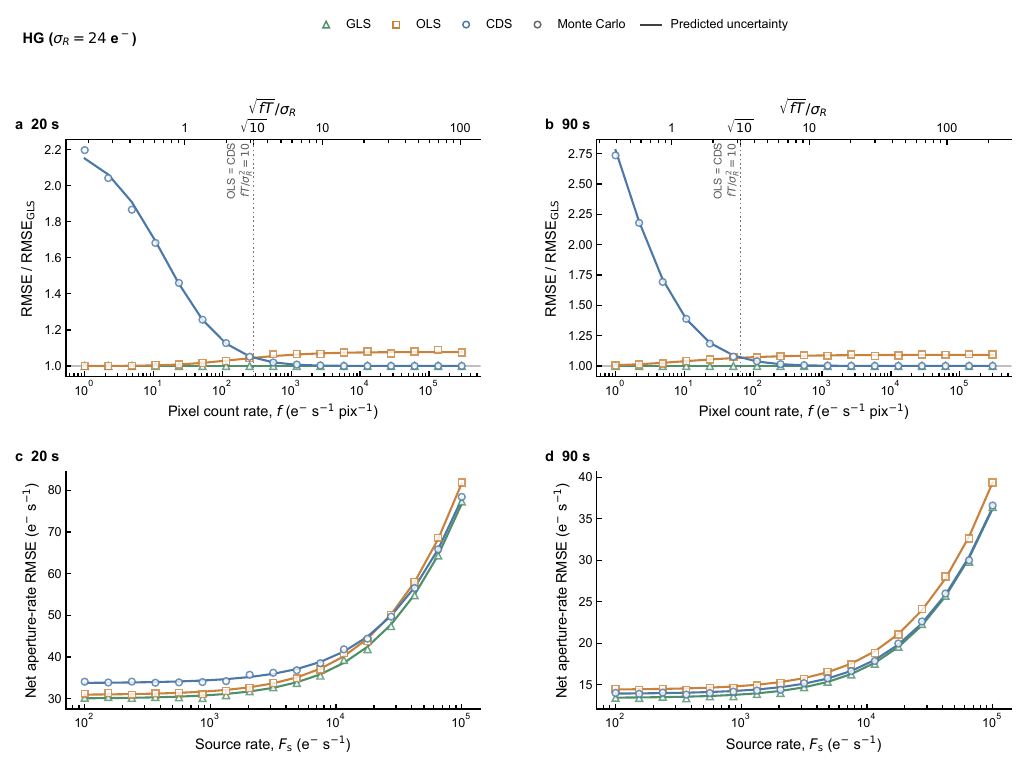}
\caption{Monte Carlo results for HG. Top: single-pixel relative root mean square error (RMSE), $\mathrm{RMSE}/\mathrm{RMSE}_\mathrm{GLS}$, versus total pixel count rate $f$. Bottom: RMSE of the net aperture count rate for a Gaussian PSF versus total source count rate $F_\mathrm{s}$, excluding sky and dark background. Continuous curves show analytical RMSE ratios in the top panels and formal aperture uncertainties in the bottom panels; open markers show Monte Carlo results. The secondary horizontal axes show $\sqrt{fT}/\sigma_\mathrm{R}$, and vertical lines mark $fT/\sigma_\mathrm{R}^2=10$.}
   \label{fig:mc-hg}
\end{figure*}

Figure~\ref{fig:mc-hg} shows the HG simulations, and Table~\ref{tab:mc-background-rmse} compares single-pixel precision penalties across the three gain modes.

The precision penalty of a simplified estimator varies with operating point. OLS approaches GLS when read noise dominates, and CDS approaches GLS when Poisson noise dominates. At fixed gain and sampling configuration, increasing the total count rate increases $\eta$, reversing the OLS--CDS RMSE ordering near $\eta=10$ (Figure~\ref{fig:mc-hg}). The crossover criterion identifies the better of the two estimators, while the curves relative to GLS quantify the corresponding precision penalty.

At the representative background operating points, OLS is the better simplified estimator for both the 20\,s and 90\,s configurations in MG and LG. Its analytical RMSE increases relative to GLS are 1.4\%/6.2\% for MG and approximately 0\%/0.24\% for LG. For HG, choosing OLS at 20\,s and CDS at 90\,s gives increases of 2.9\% and 3.7\%, respectively (Table~\ref{tab:mc-background-rmse}). These results translate the ideal ordering into configuration-specific precision penalties. All Monte Carlo results differ from the analytical predictions by less than twice the listed approximate sampling uncertainty, supporting these calculations.

\begin{table*}[!htbp]
\centering
\caption{Single-pixel relative RMSE at representative background operating points: analytical predictions and Monte Carlo simulations for the three gain modes.\label{tab:mc-background-rmse}}
\setlength{\tabcolsep}{2pt}
\scriptsize
\begin{tabular}{lccccccc}
  \hline\hline\noalign{\smallskip}
  Mode & $f$ & Configuration & $f_\mathrm{cross}$ & \multicolumn{2}{c}{OLS/GLS} & \multicolumn{2}{c}{CDS/GLS} \\
  \cline{5-6}\cline{7-8}\noalign{\smallskip}
  & (e$^-\,\mathrm{s}^{-1}\,\mathrm{pix}^{-1}$) & & (e$^-\,\mathrm{s}^{-1}\,\mathrm{pix}^{-1}$) & Theory & MC & Theory & MC \\
  \hline\noalign{\smallskip}
  HG & 113 & 20\,s & 284.4 & 1.0293 & $1.0315\pm0.0146$ & 1.1241 & $1.1247\pm0.0159$ \\
  HG & 113 & 90\,s & 65.1 & 1.0734 & $1.0730\pm0.0152$ & 1.0366 & $1.0364\pm0.0147$ \\
  MG & 120 & 20\,s & 830.1 & 1.0139 & $1.0119\pm0.0143$ & 1.3062 & $1.3407\pm0.0190$ \\
  MG & 120 & 90\,s & 189.9 & 1.0616 & $1.0619\pm0.0150$ & 1.1093 & $1.1204\pm0.0158$ \\
  LG & 109 & 20\,s & $6.259\times10^4$ & 1.0000 & $1.0000\pm0.0141$ & 2.1937 & $2.1493\pm0.0304$ \\
  LG & 109 & 90\,s & $1.432\times10^4$ & 1.0024 & $1.0033\pm0.0142$ & 3.3156 & $3.4026\pm0.0481$ \\
  \noalign{\smallskip}\hline
\end{tabular}

\tablecomments{The $\pm$ values in the MC columns are first-order approximations to the sampling uncertainty of the RMSE ratios.}
\end{table*}

In the aperture simulations, the grid median of the formal-uncertainty to empirical-standard-deviation ratio is close to unity for each configuration, supporting the propagation under independent pixels and mean background estimation. The lower panels of Figure~\ref{fig:mc-hg} show the HG net-aperture-count-rate RMSE.

\section{Observations and Analysis Methods}
\label{sect:observational_methods}

We compare the temporal repeatability of the three estimators in dark and on-sky sequences and test covariance-model predictions against the observed scatter. Integrating-sphere sequences at several illumination levels test the OLS--CDS crossover over a wider range of operating points. Unless otherwise specified, bootstrap intervals below are 95\% intervals from 5000 paired resamples, with shared resampling indices across estimators. Uncertainty in the dark and on-sky crossover times is summarized by the crossing-set envelope. Sampling units and implementation details are given below and in Appendix~\ref{sect:comparison_details}.

\subsection{Observations and Comparison Samples}
\label{sect:observations}

\noindent\textbf{Data and sampling configurations.}
The camera has a $1280\times1024$-pixel detector with $15\;\mu$m pixels, a pixel scale of $0\farcs48$\,pix$^{-1}$, a field of view of approximately $10^\prime\times8^\prime$, and a response range of $1.0$--$1.56\;\mu$m. The on-sky data comprise 44 consecutive valid ramps obtained on 2026 March 8. A separate continuous sequence of 59 dome-dark ramps provides the control data. Both sequences use the same readout configuration. The 20\,s and 90\,s products follow Table~\ref{tab:exposure-configurations}, with a nominal 0.75\,s sampling interval used throughout the fits.

To extend the observed operating-point range of the crossover test, we also use integrating-sphere sequences obtained on 2026 September 7. The environmental chamber was at $-10\,^{\circ}\mathrm{C}$ and the detector chip was maintained near 123\,K. At fixed lamp power, seven aperture settings, 00 and 02--07, were used; setting 00 had the aperture closed. Each setting contains 10 ramps with 300 reads per ramp. The HG conversion gain was 0.466\,e$^-$/ADU and the nominal sampling interval was 0.500\,s. Discarding the first and last reads leaves 298 points and a maximum fitting span of 148.5\,s. These data use the same 12 fixed regions of interest (ROIs) defined below to test relative OLS--CDS efficiency and the crossover operating points. Quality checks are described in Appendix~\ref{sect:lab_sample}.

\noindent\textbf{Count-rate products.}
Count rates in the on-sky and dark ramps are estimated using two-iteration FGLS, OLS, and CDS. FGLS initializes the ideal covariance with the median inter-read difference rate, updates it using the first fitted rate, and then performs a second fit. In this HG ramp configuration, raw reads begin to depart from a linear response above approximately 44\,000\,ADU; the calibrated nonlinearity below this limit is less than 1\%. We use this upper limit to select valid reads and apply no nonlinearity correction.

Raw on-sky ramps contain source, sky, and dark signals. Standard count-rate products are obtained by master-dark subtraction, flat-fielding, and stripe correction. The master dark is constructed from 59 dark ramps, while the flat field and stripe pattern are derived from ITR data. Random dark noise in the original on-sky exposures remains in the measurement uncertainty after dark subtraction. Background and source repeatability are evaluated using the standard count-rate products; crossover scans and covariance tests refit the raw ramps.

\begin{figure*}[!htbp]
\centering
\includegraphics[width=0.96\textwidth]{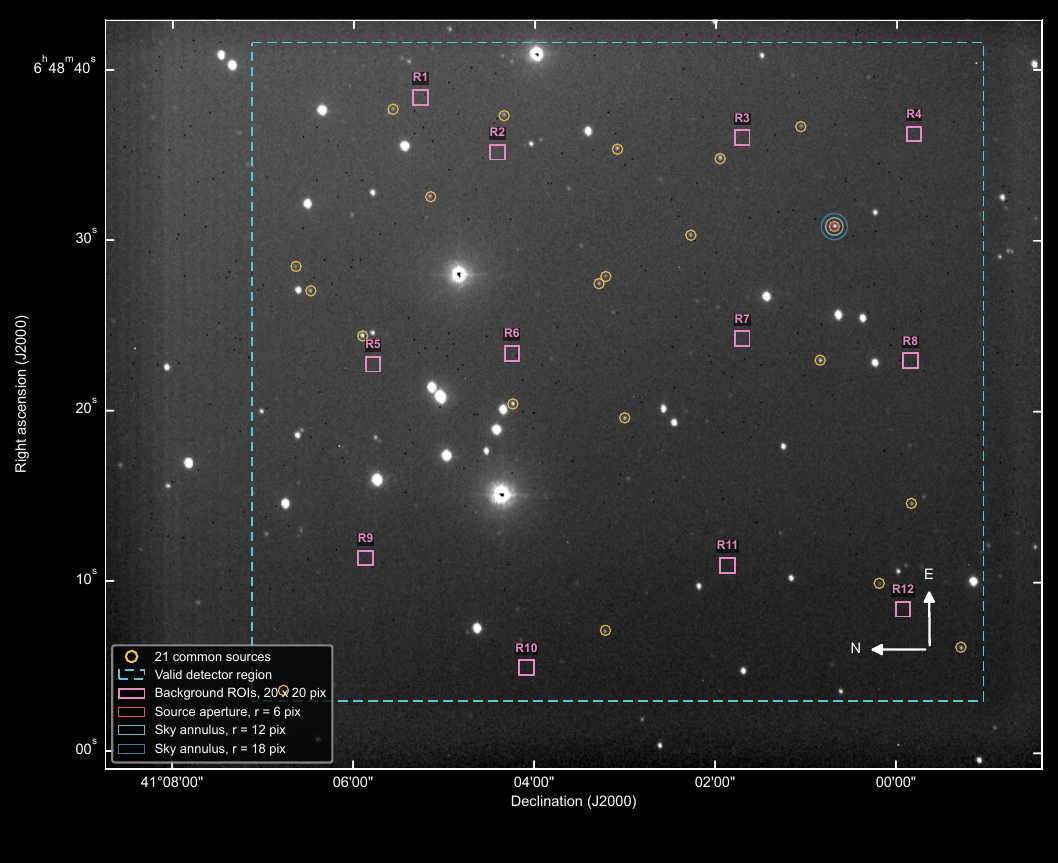}
\caption{Observed field, 21 common sources, and 12 fixed background ROIs. Yellow open circles mark all 21 common sources, magenta boxes mark the ROIs, and the cyan dashed outline bounds the valid region.}
\label{fig:onsky-field}
\end{figure*}

\noindent\textbf{Common background sample.}
To avoid column stripes on the left and nonuniform edge regions, the analysis is restricted to $200\le x<1200$ and $30\le y<930$. This region is divided into 4 columns $\times$ 3 rows, with one $20\times20$\,pix window selected per cell (Figure~\ref{fig:onsky-field}). ROIs remain fixed in native pixel coordinates across exposures. Stacked and individual-exposure detections mask sources and their trajectories. Every pixel in a candidate window must simultaneously satisfy the following criteria for all three estimators, both configurations, and all 44 on-sky exposures: finite positive count rates and formal uncertainties, the required number of valid samples, and OLS fit $r^2\ge0.7$. This procedure fixes 12 ROIs, each containing 400 pixels valid across all compared products; the dark analysis uses the same coordinates. Both background and source analyses use the OLS $r^2$ field. Threshold sensitivity is described in Appendix~\ref{sect:comparison_details}.

\subsection{Background Residual Statistics}
\label{sect:temporal_residual_methods}

Background repeatability is measured from temporal residuals after removing stable pixel differences and exposure-wide common variations. With $r$ denoting a ramp or exposure and $p$ a pixel within an ROI, we define
\begin{equation}
e_{r,p}=\hat f_{r,p}-\langle\hat f_{r,p}\rangle_r
-\langle\hat f_{r,p}\rangle_p+\langle\hat f_{r,p}\rangle_{r,p},
\label{eq:temporal_residual}
\end{equation}
These residuals have zero temporal mean for each pixel and zero spatial mean within the ROI for each exposure. For $R$ ramps or exposures and $P$ pixels valid across all compared products per ROI, the empirical scatter in each ROI is
\begin{equation}
\sigma_{\rm time}=\sqrt{\frac{\sum_{r,p}e_{r,p}^2}{(R-1)(P-1)}}.
\label{eq:temporal_residual_scatter}
\end{equation}
The overall scatter pools residual sums of squares and degrees of freedom across the 12 ROIs. The formal-error scale is the square root of the mean formal variance over the same pixels and exposures, and their ratio is $R_{\rm time}$:
\begin{equation}
\begin{gathered}
\sigma_{\rm formal}=\sqrt{\operatorname{mean}_{r,p}(\sigma_{r,p,\rm formal}^2)},
\\ R_{\rm time}=\sigma_{\rm time}/\sigma_{\rm formal}.
\end{gathered}
\label{eq:temporal_formal_scatter}
\end{equation}
For these dark temporal statistics, ramps are refitted on the nominal time grid and formal uncertainties use model M0, comprising independent white read noise and Poisson noise. Pixelwise read noise is first estimated from read differences within each ramp and then combined across ramps using a clipped mean. The total count rate is the full-sample mean OLS rate of each pixel, constrained to be nonnegative. On-sky statistics use formal uncertainties from the standard products. Appendix~\ref{sect:normalization_audit} describes the relation between shared master-dark uncertainties and temporal residuals.

The background bootstrap hierarchically resamples exposures and $5\times5$-pixel blocks within each ROI, repeating the two-way demeaning in every resample.

\subsection{OLS--CDS Crossover Analysis}
\label{sect:ols_cds_crossover_methods}

For the dark and raw on-sky data, we refit OLS and CDS to the first $n$ retained reads ($n=10,\ldots,119$) on the nominal grid, giving $T=(n-1)\times0.75$\,s. Within the 12 fixed ROIs, pooled empirical variances are calculated as in Section~\ref{sect:temporal_residual_methods} and compared with model predictions:
\begin{equation}
\begin{gathered}
Q_{\rm emp}=\frac{V_{\rm emp,OLS}}{V_{\rm emp,CDS}},\\
Q_M=\frac{V_{M,\rm OLS}}{V_{M,\rm CDS}},\\
K_{m,M}=\frac{V_{\rm emp,m}}{V_{M,m}}.
\end{gathered}
\label{eq:ols_cds_crossover_ratios}
\end{equation}
Here $m$ denotes the estimator, $M$ the covariance model, and $V$ the count-rate variance. OLS has the lower variance when $Q<1$, and CDS when $Q>1$; $K$ compares the empirical and predicted variance scales.

M0 comprises independent white read noise and Poisson noise, whereas M2 is the empirical difference covariance estimated from the training folds. Empirical curves use full-sample temporal residuals. Model parameters are estimated from training sets using the five-block partition in Section~\ref{sect:dark_cv_methods}, and fold predictions are combined with weights proportional to the number of test exposures. Both models propagate variances using the estimator weights for each read window; scan parameters and variance normalization are given in Appendices~\ref{sect:comparison_details} and~\ref{sect:normalization_audit}. For the pooled ROI curves, we define $\eta=\langle f\rangle T/\langle\sigma_\mathrm{R}^2\rangle$, weighting total count rates and read-noise variances by test-exposure and ROI-pixel counts. This aggregate coordinate corresponds to the pooled variance comparison and uses training-fold total rates and pixelwise read-noise calibration. It differs from the representative operating points in Section~\ref{sect:model_parameters}, which use the 24\,e$^-$ median read noise.

Crossover times are obtained by linearly interpolating $\log Q$ as a function of $T$ between adjacent scan points and locating intersections with $\log Q=0$. Read windows share resampling indices; Appendix~\ref{sect:comparison_details} defines the crossing-set envelope.

For the integrating-sphere sequences, OLS and CDS are fitted to prefixes of $n=4,\ldots,298$ retained reads at the nominal 0.500\,s spacing, and temporal variances are compared using Equation~\ref{eq:temporal_residual_scatter}. M0 uses the full-ramp total-rate map for each setting and an independent read-noise calibration (Appendix~\ref{sect:lab_sample}), with the same ROI-averaged definition of the operating point. Prefix scans report empirical and M0-predicted OLS/CDS variance ratios and point estimates of all crossings, without crossover intervals. Paired bootstrap intervals for the variance ratios at the full 148.5\,s span use the sampling scheme in Appendix~\ref{sect:lab_sample}. These sequences are not included in the M2 contiguous-block validation below.

\subsection{Blocked Cross-Validation of Covariance Models}
\label{sect:dark_cv_methods}

The dark and raw on-sky sequences are each divided into five contiguous blocks in acquisition order. Each block is used in turn for testing, with the remaining four used for training. Stable patterns, total count rates, and difference covariance are estimated from each training set. Dark read noise is estimated within the training fold; the on-sky analysis uses independent dark calibration. Both covariance tests use the same nominal 0.750\,s grid as the crossover scans, with identical times for slope fitting and model variance propagation. Test residuals are defined as
\begin{equation}
e^{\rm test}_{r,p}=\hat f_{r,p}-\bar f^{\rm train}_p
-\langle\hat f_{r,p}-\bar f^{\rm train}_p\rangle_p.
\label{eq:dark_heldout_residual}
\end{equation}
where $\bar f^{\rm train}_p$ is the mean count rate of pixel $p$ in the training ramps and $\langle\cdot\rangle_p$ denotes the pixel average within an ROI.

Let $\boldsymbol\Sigma_M$ denote the difference covariance. M0 corresponds to $\boldsymbol\Lambda$ in Appendix~\ref{sect:gls}; M2 pools pixelwise difference residuals from the training ramps within each ROI. On-sky M2 describes the total empirical covariance due to both detector and observational variations. M2 is used for variance propagation without changing the construction of the estimator weights. For $\hat f=\boldsymbol q^{\mathsf T}\boldsymbol d$,
\begin{equation}
V_M=\boldsymbol q^{\mathsf T}\boldsymbol\Sigma_M\boldsymbol q,
\qquad R_M=\sigma_{\rm test}/\sqrt{\langle V_{M,\rm test}\rangle}.
\label{eq:dark_covariance_propagation}
\end{equation}
The predicted variance $V_{M,\rm test}$ matches the test-residual processing, accounting for ROI common-mode subtraction, training-template estimation error, and the degrees-of-freedom correction for covariance estimation (Appendix~\ref{sect:normalization_audit}). Here $\sigma_{\rm test}=\sqrt{\operatorname{mean}_{r,p}[(e^{\rm test}_{r,p})^2]}$ is the residual RMS over the test blocks and common ROI pixels, and $\langle V_{M,\rm test}\rangle$ averages over the same set. Fixed-weight OLS is the primary test, with CDS and two-iteration FGLS as supporting comparisons. For FGLS, the fitted weights of each ramp are inserted into the propagation expression.

We adopt a scale-reference band in which the predicted standard deviation differs from the measured scatter by no more than 10\%, corresponding to $0.909\le R_M\le1.111$. Resampling holds the trained model in each fold fixed and draws only test ramps and $5\times5$-pixel blocks within each ROI. The resulting intervals are conditional on the fitted training models.

To describe temporal structure in the read differences, we define $\rho_d(\ell)=\langle(\widehat{\boldsymbol\Sigma}_2)_{j,j+\ell}\rangle_j/\langle(\widehat{\boldsymbol\Sigma}_2)_{j,j}\rangle_j$, where $\ell$ is the lag in read intervals and each average uses the corresponding valid matrix elements.

\subsection{Source Photometry and Repeatability}
\label{sect:photometry}

Sources are detected with \texttt{DAOfind} \citep{Stetson1987} at a $5\sigma$ threshold. Astrometric calibration uses \texttt{Astrometry.net} \citep{Lang2010}, followed by matching to the 2MASS catalog \citep{Skrutskie2006}.

Aperture photometry uses a radius $r=6$\,pix ($2\farcs9$) and a sky annulus with inner and outer radii of 12 and 18\,pix ($5\farcs8$--$8\farcs7$). The local background is the annular median after up to five iterations of $3\sigma$ clipping. The net aperture count rate is the sum of pixel count rates within the aperture minus the local background median multiplied by the effective aperture area. All three estimators use the same aperture geometry and are compared on exposures passing their common quality selection. The formal aperture uncertainty $\sigma_{\mathrm{ap,formal},m,s,e}$ uses the mean-background propagation approximation in Equation~\ref{eq:aperture_rate_uncertainty}, where $m$, $s$, and $e$ index estimator, source, and exposure.

The source sample requires $J>13$ to avoid bright sources with raw reads exceeding 44000\,ADU and at least 30 exposures paired across all three methods. Valid pixels must have the nominal number of usable samples and OLS $r^2\ge0.7$. Pixels within the 2\,pix source core must have finite, valid values. Valid-pixel fractions must be at least 90\% in the aperture and 80\% in the sky annulus, both of which must lie entirely within the adopted valid region. The common sample in both configurations contains 21 sources with approximately $13.70\le J\le15.18$, each retaining 32--44 paired exposures per configuration. Threshold sensitivity is described in Appendix~\ref{sect:comparison_details}.

The sources are assigned to three fixed brightness groups, from faint to bright in $J$, with seven sources per group. Let $\widetilde F_{m,s}$ be the median net aperture count rate of source $s$ across exposures. The raw absolute and fractional temporal scatter are
\begin{equation}
\begin{gathered}
  \sigma_{\mathrm{raw},m,s}=\operatorname{Std}_e(F_{m,s,e}),\\
  \sigma_{\mathrm{raw,frac},m,s}=\operatorname{Std}_e(F_{m,s,e}/\widetilde F_{m,s}-1)\,.
  \end{gathered}
\label{eq:source_raw_scatter}
\end{equation}
where $\operatorname{Std}_e$ is the sample standard deviation over paired exposures. For each estimator and exposure, a leave-one-source-out procedure excludes source $s$ and estimates the exposure common mode from the median normalized count-rate deviation of the other common sources passing quality selection in that exposure. The residual of source $s$ is then
\begin{equation}
\begin{gathered}
  c_{m,e}^{(-s)}=\operatorname{median}_{s'\ne s}\left(F_{m,s',e}/\widetilde F_{m,s'}-1\right),
  \\ r_{m,s,e}=F_{m,s,e}-\widetilde F_{m,s}-\widetilde F_{m,s}c_{m,e}^{(-s)}\,.
  \end{gathered}
\label{eq:source_common_mode}
\end{equation}
The residual scatter is $\sigma_{\mathrm{res},m,s}=\operatorname{Std}_e(r_{m,s,e})$, with fractional form $\sigma_{\mathrm{res},m,s}/\widetilde F_{m,s}$. The common mode is calculated separately for each estimator. After common-mode removal, the error-scale ratio is
\begin{equation}
\begin{gathered}
  R_{\mathrm{source},m,s}=\frac{\sigma_{\mathrm{res},m,s}}{\overline\sigma_{\mathrm{ap,formal},m,s}},\\
  \overline\sigma_{\mathrm{ap,formal},m,s}=\sqrt{\operatorname{mean}_e(\sigma_{\mathrm{ap,formal},m,s,e}^{2})}\,.
  \end{gathered}
\label{eq:source_scatter_ratio}
\end{equation}

The corresponding raw ratio is $R_{\mathrm{raw},m,s}=\sigma_{\mathrm{raw},m,s}/\overline\sigma_{\mathrm{ap,formal},m,s}$. Error-scale ratios $R$ compare empirical scatter with formal uncertainties. Relative estimator efficiency is evaluated using paired empirical scatter ratios: residual standard-deviation ratios are calculated for each source and summarized as $\exp[\operatorname{median}_s\log(\sigma_{\mathrm{res},m_1,s}/\sigma_{\mathrm{res},m_2,s})]$. Intervals hierarchically resample sources and the paired exposures of each source, sharing indices across all three estimators. Resampling retains the normalization scales and exposure common modes estimated from the original sample. As a supplementary error-model diagnostic, we define a typical formal SNR as the median over exposures of the ratio of net count rate to formal aperture uncertainty.

\section{Observational Results}
\label{sect:results}

We first quantify the precision penalties of simplified estimators using background temporal residuals, then examine the ideal selection boundary through read scans and covariance tests, and finally test how pixel-level differences carry through to common-source photometry.

\subsection{Background Precision and Error Scales}
\label{sect:temporal_estimator_results}

\begin{figure*}[!htbp]
\centering
\includegraphics[width=0.96\textwidth]{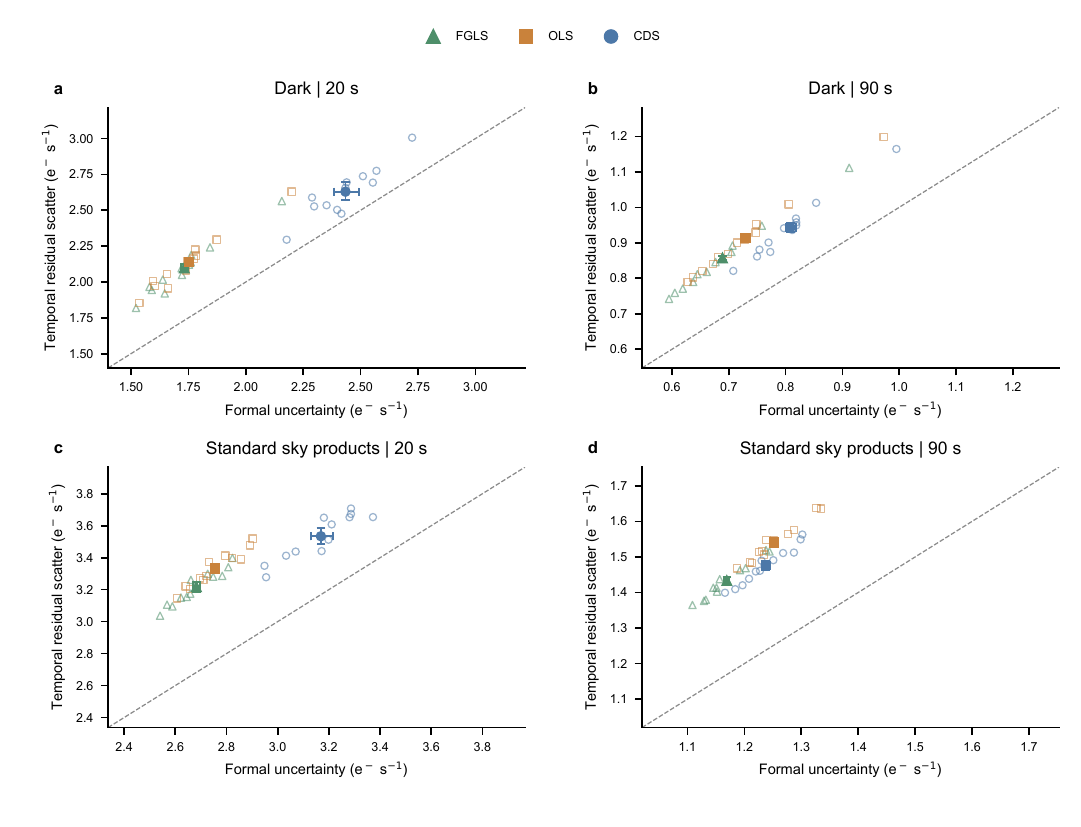}
\caption{Temporal residual scatter and formal uncertainties under definitions common to all estimators. Rows show dark and standard on-sky products; columns show the 20\,s and 90\,s configurations. Open points represent all 12 ROIs, and filled points pool sums of squares and degrees of freedom. Error bars are 95\% paired bootstrap intervals (Section~\ref{sect:temporal_residual_methods}); dashed lines indicate $y=x$.}
\label{fig:temporal-estimator-scatter}
\end{figure*}

Background temporal residuals directly quantify the precision penalty of selecting a simplified estimator. OLS has lower scatter than CDS in both dark configurations and in the 20\,s on-sky data; their ordering reverses in the 90\,s on-sky data. Two-iteration FGLS has the lowest scatter in all four data/configuration combinations (Figure~\ref{fig:temporal-estimator-scatter}). Relative to FGLS, choosing OLS at 20\,s and CDS at 90\,s increases the on-sky background standard deviation by 3.55\% and 3.01\%, respectively. The 95\% intervals of all 12 paired scatter ratios exclude unity (Table~\ref{tab:background-paired-scatter}). The ordering is unchanged with larger spatial resampling blocks, temporal blocks of two or four exposures, or omission of individual ROIs.

Paired scatter ratios measure the relative precision of background count-rate estimates, whereas empirical-to-formal ratios assess the error-model scale. All empirical-to-formal ratios have 95\% intervals above unity. Thus, even for FGLS, which has the lowest scatter, the current formal uncertainties underestimate the background temporal residual scale.

\begin{table*}[!htbp]
\centering
\caption{Paired estimator ratios for background temporal residuals. Each ratio compares the pooled residual standard deviations of the numerator and denominator methods. Brackets give 95\% paired bootstrap intervals.\label{tab:background-paired-scatter}}
\setlength{\tabcolsep}{3pt}
\small
\begin{tabular}{llccc}
\hline\hline
Data & Configuration & OLS/FGLS & CDS/FGLS & CDS/OLS \\
\hline
Dark & 20\,s & 1.0203 [1.0187, 1.0220] & 1.2543 [1.2373, 1.2746] & 1.2293 [1.2120, 1.2499] \\
Dark & 90\,s & 1.0652 [1.0626, 1.0678] & 1.1001 [1.0904, 1.1122] & 1.0328 [1.0230, 1.0446] \\
On-sky & 20\,s & 1.0355 [1.0337, 1.0372] & 1.0986 [1.0889, 1.1101] & 1.0609 [1.0511, 1.0726] \\
On-sky & 90\,s & 1.0753 [1.0724, 1.0784] & 1.0301 [1.0268, 1.0339] & 0.9579 [0.9535, 0.9628] \\
\hline
\end{tabular}

\end{table*}

\subsection{Observed OLS--CDS Crossovers}
\label{sect:ols_cds_crossover_results}

\begin{figure*}[!htbp]
\centering
\includegraphics[width=0.96\textwidth]{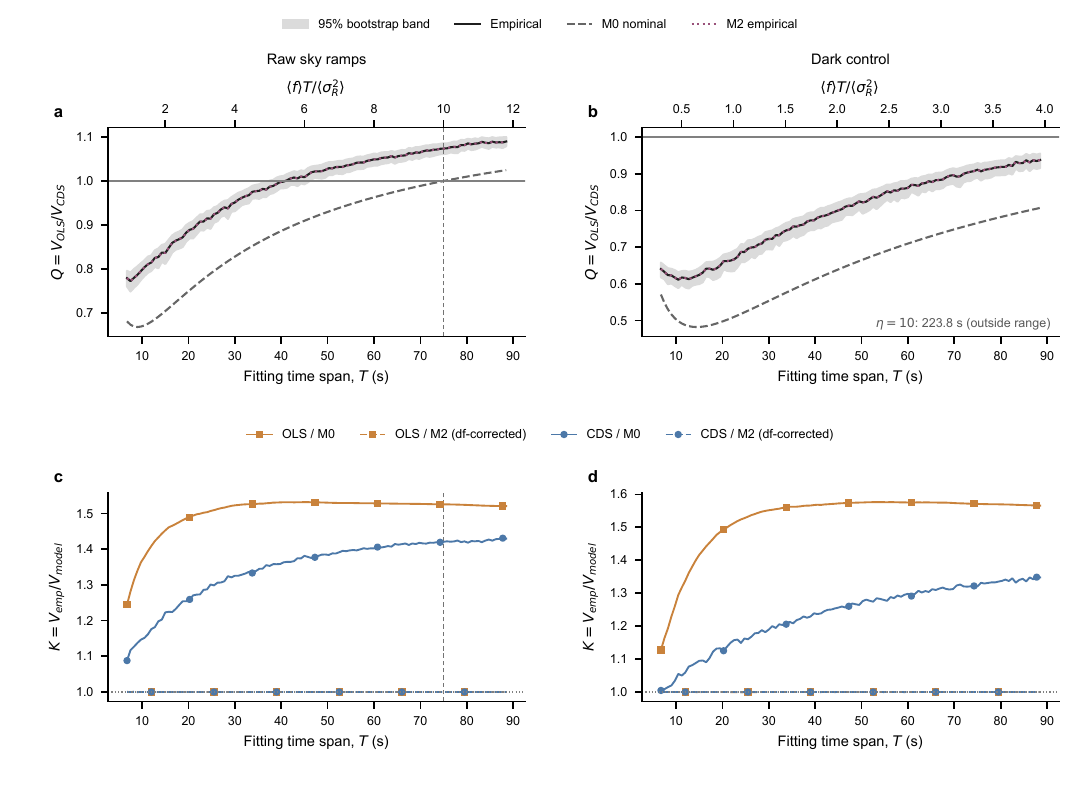}
\caption{Read-prefix scans on the nominal 0.750\,s grid. Top: OLS/CDS variance ratios for raw on-sky and dark data, with gray 95\% paired bootstrap bands. Secondary axes show $\langle f\rangle T/\langle\sigma_R^2\rangle$, using total rates and read-noise variances weighted by test-exposure and ROI-pixel counts. Vertical dashed lines mark the ideal criterion $\langle f\rangle T/\langle\sigma_R^2\rangle=10$; the dark boundary lies beyond the scan. Bottom: empirical/model variance ratios for OLS and CDS, with solid curves for M0 and dashed curves for M2 with matched degrees of freedom. Horizontal reference lines indicate unity. M2 is estimated from training folds of the same sequence.}
\label{fig:ols-cds-crossover}
\end{figure*}

The raw on-sky prefix scan increases the fitting span at fixed read spacing, increasing $\eta$ and reversing the relative efficiency of OLS and CDS. The pooled curve for the 12 ROIs crosses unity at 40.83\,s, with a bootstrap crossing-set envelope of 37.38--46.05\,s, earlier than the 75.00\,s predicted by M0 (Figure~\ref{fig:ols-cds-crossover}). M0 uses a weighted total count rate of 115.67\,e$^-\,\mathrm{s}^{-1}\,\mathrm{pix}^{-1}$ and the square root of the weighted mean read-noise variance is 29.45\,e$^-$. The earlier empirical crossover shows that the observed ordering reverses below the ideal boundary $\eta=10$.

In the full-sample scan, the M2 crossover of 40.84\,s is close to the empirical value, reflecting consistency of second-order statistics within the same sequence. Training on the first 36 on-sky ramps and testing on the final eight gives an empirical test-set crossover of 39.96\,s, compared with M2 and M0 predictions of 41.01\,s and 74.94\,s. This final-block test supports the empirical covariance description of the earlier crossover within the same sequence (Appendix~\ref{sect:normalization_audit}).

OLS has lower variance throughout the dark scan. On-sky checks using different window positions and training-rate estimators retain the earlier-crossover result.

\begin{figure*}[!htbp]
\centering
\includegraphics[width=0.96\textwidth]{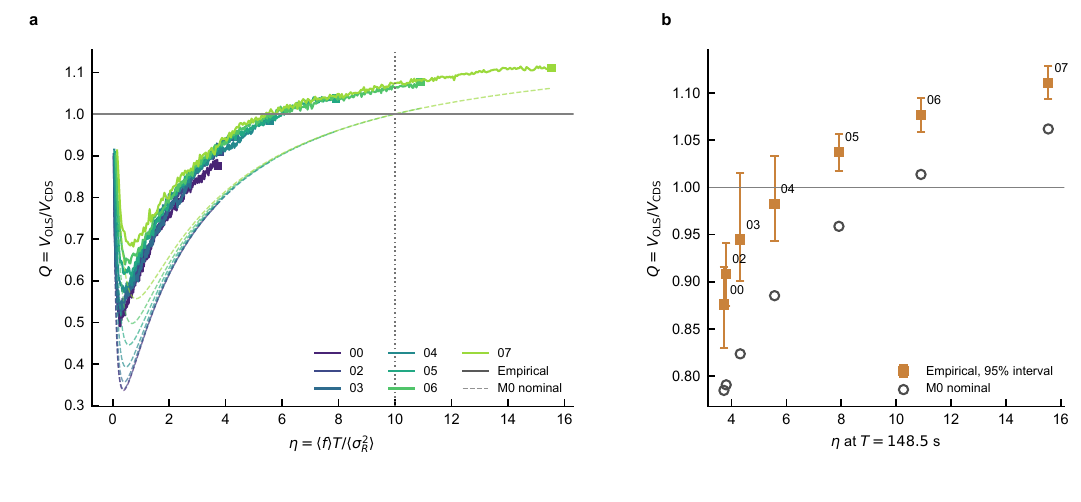}
\caption{OLS--CDS operating-point tests for seven integrating-sphere aperture settings. Each setting contains 10 ramps and uses the same 12 fixed ROIs as the on-sky analysis. Left: prefix scans from 4 to 298 reads, with solid empirical variance ratios and dashed M0 predictions at the corresponding finite sample sizes. Squares mark the full-span endpoints; the vertical dotted line marks $\eta=10$. Prefixes share reads, and no confidence bands are shown. Right: empirical ratios at 148.5\,s with 95\% percentile intervals from 5000 paired bootstrap resamples; open circles show M0. The horizontal line $Q=1$ indicates equal variance, with lower OLS scatter at $Q<1$. Setting numbers are identifiers only; the horizontal coordinate uses measured total rates and fixed calibrated read noise.}
\label{fig:lamp-crossover}
\end{figure*}

The integrating-sphere sequences extend the comparison to a wider operating-point range. Mean total count rates across the seven settings span 21.43--89.58\,e$^-\,\mathrm{s}^{-1}\,\mathrm{pix}^{-1}$, corresponding to $\eta=3.72$--15.55 at the fixed 148.5\,s span. Ordered by increasing measured total rate, the empirical variance ratio $Q=V_{\rm OLS}/V_{\rm CDS}$ rises from 0.876 to 1.111, indicating a transition from lower OLS variance to lower CDS variance (Figure~\ref{fig:lamp-crossover}). At setting 05, with $\eta=7.91$, M0 predicts $Q=0.959$, whereas the measured ratio is 1.037 with a 95\% paired bootstrap interval of [1.018,1.056]. All crossings in the prefix scans of settings 05--07 occur at $\eta\simeq5.3$--6.1, also below the ideal boundary of 10. This range summarizes point estimates of multiple crossings within the transition region. Applying common pixel-quality conditions preserves both the full-span ordering and the earlier crossovers (Appendix~\ref{sect:lab_sample}).

\subsection{Covariance Predictions of Residual Scatter}
\label{sect:dark_cv_results}

Contiguous-block cross-validation further tests how the covariance models describe OLS test-residual scales. M0 underpredicts the OLS residual scale for both dark and raw on-sky data. With training-fold empirical covariance M2, empirical-to-predicted standard-deviation ratios approach unity, and all 95\% conditional intervals fall within the predefined scale-reference band (Figure~\ref{fig:dark-blocked-cv}). Variance propagation and normalization are given in Appendix~\ref{sect:normalization_audit}.

\begin{figure*}[!htbp]
\centering
\includegraphics[width=0.96\textwidth]{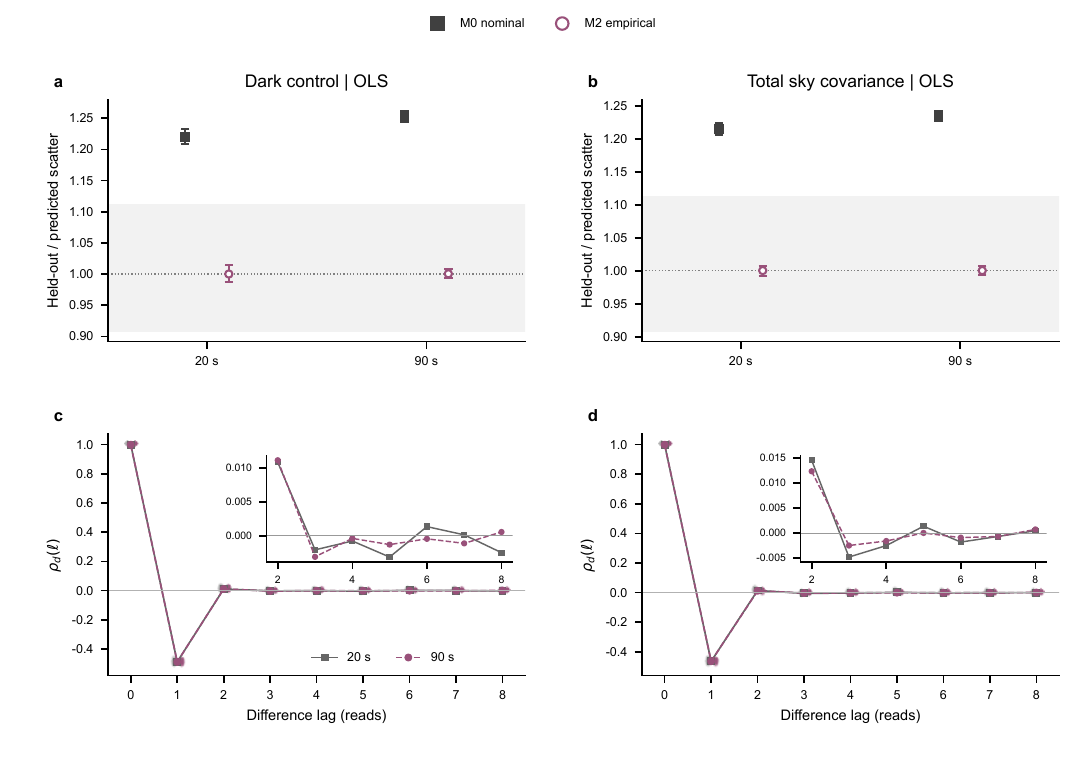}
\caption{Contiguous-block cross-validation of the covariance models. Top: OLS test-residual RMS divided by predicted standard deviation. Squares show M0, open circles M2, and error bars 95\% conditional bootstrap intervals. The gray band is the $0.909$--$1.111$ scale reference. Bottom: normalized lag covariance of read differences, $\rho_d(\ell)$, for lags 0--8; lag 0 equals unity by definition. Faint points show all 12 ROIs and five folds. Curves are arithmetic means over ROIs and folds for each of the 20\,s and 90\,s configurations; insets enlarge the mean values at lags 2--8. The dark data provide a control sequence, whereas the on-sky covariance includes the total raw variations.}
\label{fig:dark-blocked-cv}
\end{figure*}

Both data sets show positive ROI/fold-averaged covariance at lag 2 (Figure~\ref{fig:dark-blocked-cv}), indicating temporal structure beyond independent white read noise and Poisson noise.

\subsection{Photometric Repeatability and Error Scales}
\label{sect:source_temporal_scatter}

The source photometry does not clearly resolve most differences between estimators: most paired scatter-ratio intervals include unity. In Table~\ref{tab:source-paired-scatter}, only the 20\,s CDS/FGLS ratio, 1.097 [1.004, 1.187], has a 95\% interval excluding unity; the other five include it. Relaxing the $r^2$ threshold changes this point estimate only slightly, but whether its interval includes unity depends on the threshold (Appendix~\ref{sect:comparison_details}).

\begin{table*}[!htbp]
\centering
\caption{Paired residual standard-deviation ratios for the 21 common sources after common-mode removal.\label{tab:source-paired-scatter}}

\small
\begin{tabular}{llc}
\hline\hline
Configuration & Estimator Ratio & Scatter Ratio (95\% Interval) \\
\hline
20\,s & OLS/FGLS & 1.040 [0.986, 1.109] \\
20\,s & CDS/FGLS & 1.097 [1.004, 1.187] \\
20\,s & CDS/OLS & 1.038 [0.944, 1.155] \\
90\,s & OLS/FGLS & 1.019 [0.959, 1.103] \\
90\,s & CDS/FGLS & 1.019 [0.963, 1.067] \\
90\,s & CDS/OLS & 0.994 [0.913, 1.060] \\
\hline
\end{tabular}

\tablecomments{Point estimates are the exponentiated medians of the logarithmic source-by-source scatter ratios. Intervals use 5000 shared source/exposure resamples.}
\end{table*}

\begin{figure*}[!htbp]
   \centering
   \includegraphics[width=0.96\textwidth]{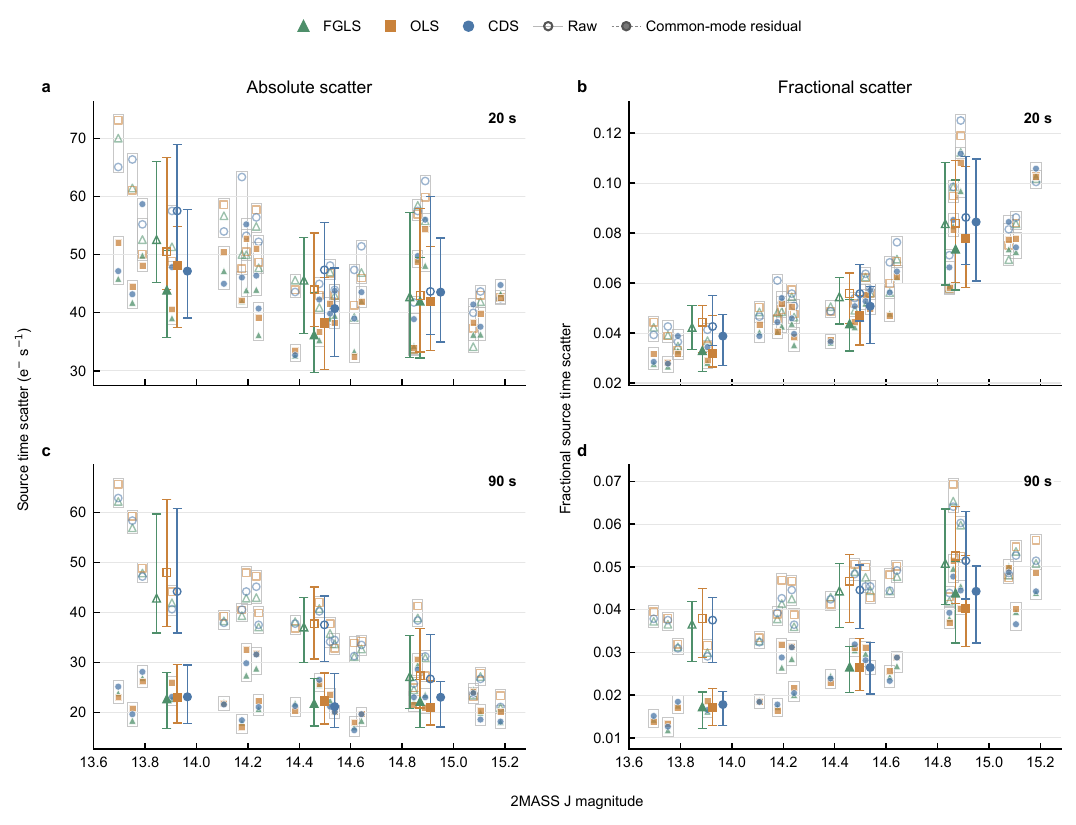}
\caption{End-to-end temporal repeatability by source brightness. Rows show the 20\,s and 90\,s configurations; columns show absolute and fractional scatter versus 2MASS $J$ magnitude. Open symbols represent raw scatter, and filled symbols represent residual scatter after leave-one-source-out removal of the exposure common mode. Colors and symbols distinguish FGLS, OLS, and CDS. Gray boxes connect the three estimators for each source. Large points and error bars show medians and 95\% paired bootstrap intervals for the fixed brightness groups.}
   \label{fig:source-temporal-scatter}
\end{figure*}

The 21 common sources have better photometric repeatability in the 90\,s configuration than in the 20\,s configuration, with further improvement after exposure common-mode removal (Figure~\ref{fig:source-temporal-scatter}). The absolute residual scatter is approximately $47$--$51\%$ lower at 90\,s, and the fractional scatter also decreases. Point estimates across brightness groups do not show a consistent monotonic ordering in both configurations.

\begin{figure*}[!htbp]
   \centering
   \includegraphics[width=0.96\textwidth]{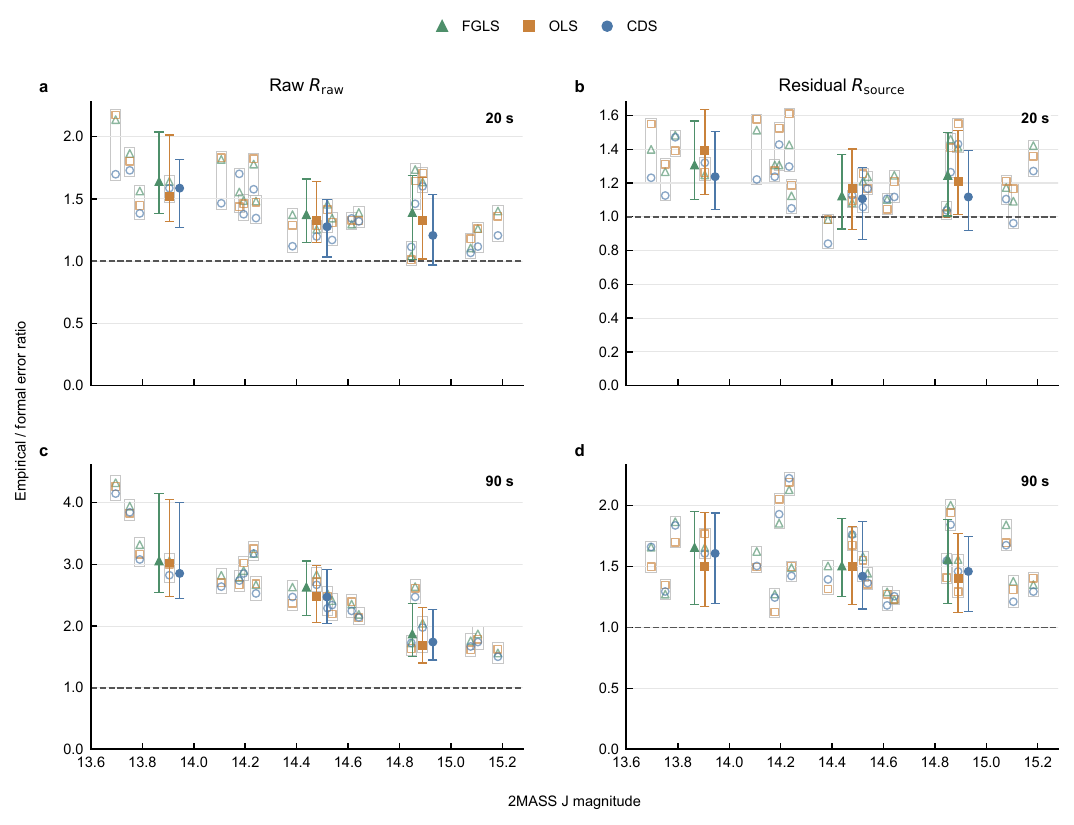}
\caption{Ratios of source temporal scatter to formal aperture uncertainty. Rows show the 20\,s and 90\,s configurations. Columns show $R_{\rm raw}$, retaining common exposure variations, and $R_{\rm source}$ after common-mode removal. Dashed lines indicate unity. Gray boxes connect FGLS, OLS, and CDS measurements of each source. Large points and error bars show medians and 95\% paired bootstrap intervals for the three fixed $J$-magnitude groups.}
   \label{fig:source-error-ratio}
\end{figure*}

Residual scatter remains above the formal aperture uncertainties (Figure~\ref{fig:source-error-ratio}). Across estimators, the all-source median $R_{\rm source}$ ranges from $1.17$--$1.26$ at 20\,s and $1.46$--$1.56$ at 90\,s; the 95\% intervals for all three methods in both configurations lie above unity. Although the measured scatter is smaller at 90\,s, the formal uncertainties decrease faster, producing a larger empirical-to-formal ratio. Formal SNR follows the error-model ordering: FGLS is highest in both configurations; between OLS and CDS, OLS is favored at 20\,s and CDS at 90\,s.

\section{Discussion}
\label{sect:discussion}

These results support separating estimator selection into assessment of the operating point and evaluation of the observed precision penalty. The ideal criterion orders the OLS and CDS candidates, background data test that ordering and the loss relative to FGLS, and source photometry tests the importance of these differences in the final measurement. We discuss the selection boundary, the different measurement levels, practical choices, and the scope of the results and future validation.

\subsection{Temporal Covariance and Estimator Efficiency}
\label{sect:temporal_covariance}

The ideal criterion provides a baseline boundary for choosing between OLS and CDS. The observed on-sky crossover occurs earlier, and empirical temporal covariance describes this shift. Count-rate variance is determined by $\boldsymbol q^{\mathsf T}\boldsymbol\Sigma\boldsymbol q$. Because OLS and CDS weight the reads differently, they respond differently to the same covariance structure. The M2 description of both the earlier crossover and the test-set error scale shows that temporal correlations affect the absolute formal-error scale as well as relative estimator efficiency (Sections~\ref{sect:ols_cds_crossover_results} and~\ref{sect:dark_cv_results}). The effects of correlated read noise on photometric errors have also been considered in other near-infrared time series \citep{Schlawin2020}.

The integrating-sphere measurements show earlier crossovers at a different sampling configuration and several illumination levels (Section~\ref{sect:ols_cds_crossover_results}), further illustrating the dependence of the practical selection boundary on the noise structure. We treat the additional temporal correlations as measured noise characteristics of the present camera state, evaluate OLS and CDS using residual comparisons, and use empirical temporal covariance to describe their precision differences and crossover locations.

FGLS weights based on an approximate covariance reduce the observed scatter in the present background data, yet their formal uncertainties remain below the measured temporal scatter. Estimator efficiency and error calibration are therefore related but distinct: paired empirical scatter ratios assess the former, while agreement between predicted and measured error scales tests the latter.

\subsection{From Background Precision to Source Photometry}
\label{sect:background_to_sources}

Background ROI statistics remove stable pixel terms and local exposure common modes. Source aperture photometry combines multiple pixels and a sky annulus and is also affected by transparency, PSF, and centroid variations. These contributions can obscure the smaller estimator differences measured in background pixels.

For the present source sample, the improvement from 20\,s to 90\,s and the improvement after exposure common-mode removal are more clearly established than the differences between estimators. Formal SNR gives the ordering expected from the error model, but most estimator differences are not clearly resolved by the source photometry (Table~\ref{tab:source-paired-scatter}). Estimator selection for source photometry should therefore be assessed using the temporal repeatability of the same sources after aperture measurement and calibration.

\subsection{Practical Estimator Selection}
\label{sect:gain_guidance}

A practical procedure begins by calculating $\eta$ from the single-read noise, total count rate, and fitting span in the relevant gain mode. The ideal boundary $\eta=10$ identifies OLS or CDS as a candidate, and its precision loss relative to GLS at the corresponding sampling configuration indicates its suitability. Paired scatter comparisons in representative observations then test and, if necessary, revise the candidate ordering. The loss of the better method relative to FGLS can subsequently be assessed against the measurement requirements. Table~\ref{tab:estimator-selection} places these two forms of evidence side by side, using the camera's representative 20\,s and 90\,s configurations.

\begin{table*}[!htbp]
\centering
\caption{Simplified-estimator choices and precision penalties at the operating points examined for this camera.\label{tab:estimator-selection}}

\setlength{\tabcolsep}{3pt}
\small
\begin{tabular}{>{\raggedright\arraybackslash}p{.23\textwidth}>{\raggedright\arraybackslash}p{.12\textwidth}>{\raggedright\arraybackslash}p{.32\textwidth}>{\raggedright\arraybackslash}p{.25\textwidth}}
\hline\hline
Operating Condition & Preferred Estimator & Precision Penalty & Evidence \\
\hline
HG dark, 20/90\,s & OLS & Standard deviation 2.03\%/6.52\% above FGLS & Dark temporal residuals \\
HG on-sky background, 20\,s & OLS & Standard deviation 3.55\% above FGLS & Standard-product temporal residuals \\
HG on-sky background, 90\,s & CDS & Standard deviation 3.01\% above FGLS & Standard-product temporal residuals \\
HG representative background, 20/90\,s & OLS/CDS & RMSE 2.9\%/3.7\% above GLS & Ideal model and MC \\
MG representative background, 20/90\,s & OLS & RMSE 1.4\%/6.2\% above GLS & Ideal model and MC \\
LG representative background, 20/90\,s & OLS & RMSE approximately 0\%/0.24\% above GLS & Ideal model and MC \\
\hline
\end{tabular}

\tablecomments{Observed rows use the point estimates in Table~\ref{tab:background-paired-scatter}, with increases defined as $100(\sigma_m/\sigma_{\rm FGLS}-1)$; paired intervals are given there. Theoretical rows use the representative parameters in Table~\ref{tab:mc-background-rmse} and GLS with known covariance. The two forms of evidence evaluate observed scatter and ideal-model RMSE, respectively.}
\end{table*}

In the present on-sky background, the approximately 3\%--4\% increase in standard deviation of the better simplified estimator relative to FGLS provides a quantitative trade-off; its acceptability depends on the required measurement precision. If other errors are approximately independent and do not change with estimator, the smaller the pixel-estimation contribution to the total variance, the smaller the impact of this increase on the final measurement. Simpler implementation motivates this work, but actual computational gains require implementation-level performance benchmarks.

Calibration parameters and representative observations can jointly establish the selection rule. It can be reused within a validated operating range while the readout configuration and noise state remain stable, with reassessment after substantial changes in conditions.

\subsection{Scope and Future Validation}
\label{sect:limitations}

The analytical criterion and limiting variance ratios apply to the constant-flux, uniform-sampling model with independent white read noise and Poisson noise in Section~\ref{sect:theory}. MG/LG choices are supported by models and simulations. On-sky background and source-photometry results come from a single night of HG observations; dark data provide a control sequence, and the integrating-sphere data extend operating-point comparisons to another configuration. The M2 empirical-covariance description applies to comparisons of error scales and crossover locations within the same sequence.

Identifying and suppressing the physical sources of the additional temporal correlations are priorities for future camera and readout-system improvements. Comparing noise amplitudes, temporal covariance, and OLS--CDS crossover locations before and after modifications would assess noise suppression and establish estimator choices at the resulting operating points.

Observations of the same field on independent nights, accompanied by contemporaneous dark sequences and supplemented by medium- and low-gain data, are needed to test the stability and scope of applicability of the selection rule. Individual read times, temperatures, and readout configurations should be recorded to compare camera states. The effects of within-ramp flux variations on uncertainty propagation and the bias of unequally weighted estimators also require testing \citep{Li2026}. Camera-specific interpixel-capacitance (IPC) and spatial-covariance calibration could further improve the source aperture-error model.

\section{Conclusions}
\label{sect:conclusions}

Using GLS as a precision benchmark, we have compared the applicability and precision penalties of OLS and CDS across noise operating points of an InGaAs camera. For constant flux, uniform sampling, and independent white read noise plus Poisson noise, established finite-sampling variance expressions give equal OLS and CDS variances at $\eta=fT/\sigma_\mathrm{R}^2=10$ for $N\ge4$. OLS has lower variance for $\eta<10$, and CDS for $\eta>10$. OLS approaches GLS when read noise dominates, and CDS approaches GLS when Poisson noise dominates. The crossover criterion thus determines the relative ordering of the two simplified estimators, while their precision loss relative to GLS determines whether they offer a comparably precise alternative.

At the representative background operating points examined here, theory and simulations favor OLS in both the 20\,s and 90\,s configurations for medium and low gain, and favor OLS and CDS, respectively, for high gain. Their losses relative to GLS are listed in Table~\ref{tab:estimator-selection}. In the HG on-sky data, choosing OLS at 20\,s and CDS at 90\,s gives background residual standard deviations 3.55\% and 3.01\% above two-iteration FGLS, quantifying the observed precision penalties in these configurations. Medium- and low-gain choices rely on theory and simulations using calibrated parameters; observational validation is limited to high gain. For the 21 common sources, the present photometry does not clearly resolve most estimator differences.

Actual temporal covariance also shifts the OLS--CDS selection boundary. The empirical on-sky crossover is 40.83\,s, earlier than the ideal prediction of 75.00\,s. Empirical temporal covariance describes the earlier crossover and residual scale within the same sequence, and integrating-sphere comparisons at several illumination levels also show earlier crossovers. The ideal criterion can therefore guide initial candidate selection, while the practical crossover must be checked against the temporal covariance of the relevant configuration.

OLS and CDS can provide simpler alternatives with precision close to GLS at appropriate operating points. The parameter $\eta$ determines their ideal relative advantage; their precision loss relative to GLS or FGLS, together with the contribution of pixel count-rate estimation to the final error budget, determines the value of substitution. The selection boundary must then be checked against the measured temporal covariance.

\normalem
\begin{acknowledgments}
This work was supported by the National Key R\&D Program of China (Grant Nos. 2024YFA1612000 and 2024YFA1612003), the Yunnan Key Laboratory of Solar Physics and Space Science (Grant No. 202205AG070009), and the Yunnan Revitalization Talent Support Program (Grant Nos. 202305AS350029 and 202305AT350005).
We extend our special thanks to Prof. Kai-Fan Ji.
We thank the Kunming Institute of Physics for supplying the infrared detector chips for our tests, and Sun Yat-sen University for providing the telescope platform that enabled our observations. We are sincerely grateful to all individuals who provided assistance, expertise, and encouragement during this work.
Generative AI tools (GPT-5.6 Sol, GPT-6 Astra, Kimi K2.6, and Kimi K3) assisted with English translation, language polishing, LaTeX formatting, and the drafting and revision of analysis and plotting code. The authors reviewed and validated all AI-assisted outputs and take full responsibility for this work.
\end{acknowledgments}

\software{Astrometry.net \citep{Lang2010}, Astropy \citep{Astropy2022}, Matplotlib \citep{Hunter2007}, NumPy \citep{Harris2020}, pandas \citep{McKinney2010}}

\section*{Data Availability}
The supporting data and analysis code are publicly available in Science Data Bank (ScienceDB) at DOI \href{https://doi.org/10.57760/sciencedb.014x2}{10.57760/sciencedb.014x2} and CSTR \href{https://cstr.cn/31253.11.sciencedb.014x2}{31253.11.sciencedb.014x2}.

\appendix
\renewcommand{\theHequation}{appendix.\thesection.\arabic{equation}}
\section{Noise Model and Count-Rate Estimators}
\label{sect:estimator_details}

\subsection{Signal and Noise Model}
\label{sect:signal_model}

Using the notation of Section~\ref{sect:theory}, uniformly sampled reads are
\begin{equation}
  y_i = b + f\,t_i + n_i \,,
  \label{eq:signal_model}
\end{equation}
where $b$ is a fixed offset, $t_i$ is time since reset, and $n_i$ contains independent white read noise and accumulated Poisson fluctuations, with covariance \citep{Rauscher2007}
\begin{equation}
  C_{ij} \equiv \langle n_i\, n_j \rangle = \sigma_\mathrm{R}^2\,\delta_{ij} + f\,\min(t_i, t_j)\,,
  \label{eq:covariance}
\end{equation}
Here $\delta_{ij}$ is the Kronecker delta. For fixed linear weights, $\hat f=\mathbf w^{\mathsf T}\mathbf y$ and $\operatorname{Var}(\hat f)=\mathbf w^{\mathsf T}\mathbf C\mathbf w$; unbiasedness requires $\sum_iw_i=0$ and $\sum_iw_it_i=1$.

\subsection{CDS and OLS}
\label{sect:cds}

The CDS estimator and its variance are
\begin{equation}
  \hat{f}_\mathrm{CDS} = \frac{y_N - y_1}{T}\,,
  \label{eq:cds_estimator}
\end{equation}
\begin{equation}
\mathrm{Var}(\hat f_\mathrm{CDS})=\frac{2\sigma_\mathrm R^2+fT}{T^2}.
\label{eq:cds_variance}
\end{equation}
OLS fits a line to all reads with equally weighted residuals, yielding
\begin{equation}
  \hat{f}_\mathrm{OLS} = \frac{\sum_{i=1}^{N}(t_i - \bar{t})(y_i - \bar{y})}{\sum_{i=1}^{N}(t_i - \bar{t})^2}\,,
  \label{eq:ols_estimator}
\end{equation}
where $\bar t=N^{-1}\sum_i t_i$ and $\bar y=N^{-1}\sum_i y_i$. Equal weighting refers to the fit residuals; the weights in the linear combination for the slope are $(t_i-\bar t)/\sum_j(t_j-\bar t)^2$. For uniform sampling, the exact variance is
\begin{equation}
  \mathrm{Var}(\hat{f}_\mathrm{OLS}) = \frac{12\,\sigma_\mathrm{R}^2}{\Delta t^2\,N(N^2-1)} + \frac{6\,f\,(N^2+1)}{5\,\Delta t\,N(N^2-1)}\,.
  \label{eq:ols_variance_exact}
\end{equation}
This expression follows from Equation~(1) of \citet{Rauscher2007} with one read per group, dividing the integrated-signal variance by $T^2$. Table~\ref{tab:limiting-variance-ratios} gives the two noise limits.

\subsection{GLS in Difference Space}
\label{sect:gls}

Let $d_j=y_{j+1}-y_j$ for $j=1,\ldots,N-1$, so that $\mathbf d=f\Delta t\mathbf 1+\boldsymbol\varepsilon$. The difference covariance is
\begin{equation}
  \Lambda_{jk} = \begin{cases}
    2\sigma_\mathrm{R}^2 + f\,\Delta t & j = k \,, \\
    -\sigma_\mathrm{R}^2                & |j - k| = 1 \,, \\
    0                                    & |j - k| > 1 \,.
  \end{cases}\,.
  \label{eq:lambda_tridiag}
\end{equation}
For known covariance, the GLS estimator and variance are \citep{Brandt2024}
\begin{equation}
  \hat{f}_\mathrm{GLS} = \frac{\mathbf{1}^\mathsf{T}\boldsymbol{\Lambda}^{-1}\mathbf{d}}{\Delta t\;\mathbf{1}^\mathsf{T}\boldsymbol{\Lambda}^{-1}\mathbf{1}}\,,
  \label{eq:gls_estimator}
\end{equation}
\begin{equation}
  \mathrm{Var}(\hat{f}_\mathrm{GLS}) = \frac{1}{\Delta t^2\;\mathbf{1}^\mathsf{T}\boldsymbol{\Lambda}^{-1}\mathbf{1}}\,.
  \label{eq:gls_variance}
\end{equation}
In practice, the weights are obtained by solving the tridiagonal system $\boldsymbol\Lambda\boldsymbol z=\boldsymbol1$, giving $\hat f=\boldsymbol z^{\mathsf T}\boldsymbol d/(\Delta t\,\boldsymbol1^{\mathsf T}\boldsymbol z)$ with a per-pixel computational cost of $\mathcal O(N)$ \citep{Brandt2024}.

\section{Comparison Samples and Statistical Conventions}
\label{sect:comparison_details}

\subsection{Common Samples and Selection Sensitivity}

Lowering the OLS goodness-of-fit threshold from $r^2\ge0.7$ to 0.5, or requiring only finite values, leaves the statistical inputs in the 12 fixed background ROIs unchanged while slightly increasing the number of paired source exposures; more pixels and candidate windows also become eligible within the valid region. For the fixed sample of 21 sources, the 20\,s CDS/FGLS residual-scatter ratio changes little in point estimate, but inclusion of unity in its 95\% interval depends on the threshold.

Sensitivity of the background estimator ratios to short-range exposure correlations is assessed with circular moving-block resampling, using block lengths of one, two, and four exposures and 5000 resamples each. Spatial blocks remain $5\times5$ pixels. All three estimators share resampling indices, and two-way demeaning is repeated for every resample. All 12 estimator-ratio intervals exclude unity at each temporal block length.

\subsection{Scan Parameters and Interval Definitions}

In the scans, M0 uses a fixed read-noise map and 119-point OLS rate maps from each training fold. Rates are combined pixelwise with a $3\sigma$-clipped mean and constrained to be nonnegative. Total rates and read-noise variances are averaged within ROIs, and folds are weighted by their numbers of test ramps. On-sky total rates include sky and dark signals. M2 is constructed from training-fold 119-point read differences, after which the appropriate window submatrices are extracted. Variance normalization is described in Appendix~\ref{sect:normalization_audit}.

The scan's 5000 paired bootstrap resamples draw full-sample squared-residual contributions, sharing exposures and $5\times5$-pixel blocks across all $n$ and between OLS and CDS. Each variance distribution is recentered so that its median equals the point estimate. By contrast, the background temporal statistics in Section~\ref{sect:temporal_residual_methods} repeat two-way demeaning in every resample. All crossings of each resampled scan curve are retained. The lower and upper bounds of the crossing-set envelope are the 2.5th percentile of the earliest crossings and the 97.5th percentile of the latest crossings, respectively.

\subsection{Integrating-sphere Sample and Resampling}
\label{sect:lab_sample}

The integrating-sphere sample uses seven aperture settings, 00 and 02--07, with 00 closed. The nonnegative pixelwise mean total rate from each setting's full OLS ramps is held fixed for all prefixes. It includes the dark signal and background of that experiment; setting 00 is not subtracted. The independent read-noise calibration has an RMS of 29.25\,e$^-$ within the ROIs.

Full-span intervals bootstrap the 10 ramps within each setting and $5\times5$-pixel blocks within each ROI, repeating two-way demeaning in every resample. OLS and CDS share resampling indices; ramps are resampled independently by setting, while spatial blocks are shared across settings. We report percentile intervals without recentering. Prefixes share reads, and the scans report only point estimates of all crossings.

Excluding pixels with high reads or zero codes preserves the full-span estimator ordering and the earlier crossovers at the higher illumination settings.

\section{Residual Variance Propagation and Shared Calibration Terms}
\label{sect:normalization_audit}

Let the training set contain $n_{\rm tr}$ ramps and each ROI contain $P=400$ pixels. Stable pixel templates and the ROI common component of each ramp are removed from the training differences, leaving residual vectors $\boldsymbol u_{r,p}$. Accounting for the degrees of freedom used by the training mean, we define
\begin{equation}
\widehat{\boldsymbol\Sigma}_{2,\rm proj}
=\frac{\sum_{r,p}\boldsymbol u_{r,p}\boldsymbol u_{r,p}^{\mathsf T}}{(n_{\rm tr}-1)P}.
\end{equation}
For fixed-weight OLS/CDS, this gives \[
\boldsymbol q^{\mathsf T}\widehat{\boldsymbol\Sigma}_{2,\rm proj}\boldsymbol q=\sum_{r,p}(\boldsymbol q^{\mathsf T}\boldsymbol u_{r,p})^2/[(n_{\rm tr}-1)P].
\] Covariance propagation is therefore equivalent to the spatial mean of the per-pixel sample variances of training slope residuals. This algebraic identity does not require independent ramps. If training and test ramps are independent and identically distributed, estimating the template from a finite training set inflates the test-residual variance by $(1+1/n_{\rm tr})$, giving the M2 prediction
\begin{equation}
V_{2,\rm test}=(1+1/n_{\rm tr})
\boldsymbol q^{\mathsf T}\widehat{\boldsymbol\Sigma}_{2,\rm proj}\boldsymbol q.
\end{equation}
For the M0 pixelwise slope variance $v_p$, the diagonal term after spatial projection is $v_p(1-2/P)+\sum_{p'}v_{p'}/P^2$, followed by the same template factor. M2 describes the mean temporal covariance after ROI common-mode removal; propagation with data-dependent FGLS weights remains approximate.

The final-block crossover test trains on the first 36 on-sky ramps and tests on the last eight, constructing OLS and CDS weights for prefixes $n=10,\ldots,119$. The empirical quantity is the test-residual mean square from Equation~\ref{eq:dark_heldout_residual}. M0 and M2 use only training data and the fixed read-noise calibration, matching both spatial projection and the training-template factor. After equally weighting the 12 ROIs, crossings are all intersections of the estimator mean-square ratio with unity. Each of the three curves has one crossing in this test, reported as a point estimate.

The full-sample scan compares the degrees-of-freedom-corrected statistic $\sum e^2/[(R-1)(P-1)]$. It therefore uses $P/(P-1)$ times $\widehat{\boldsymbol\Sigma}_{2,\rm proj}$, without the test-template factor.

Formal uncertainties in the standard products include a master-dark calibration term. In the present background statistics, the ratio of its ROI-averaged variance to the mean product formal variance ranges from 0.062--0.838\% across the 12 ROIs, both configurations, and all three estimators. Temporal demeaning at fixed detector coordinates cancels the fixed error of the shared master dark, while dark-signal Poisson noise in the original exposures remains part of the temporal variance. Calibration uncertainties for moving-source aperture photometry require separate propagation.

\bibliographystyle{aasjournalv7.1}
\bibliography{references}

\end{document}